# Intrinsically single-mode vertical cavities: towards PT-symmetric surface-emitting lasers

H. F. J̲ᴏɴᴇꜱ[1], Mʏᴋᴏʟᴀ Kᴜʟɪꜱʜᴏᴠ[2],*

[1] *Physics Department, Imperial College, London, SW7 2BZ, UK*
[2] *HTA Photomask, 1605 Remuda Lane, San Jose, CA, 95112, USA*
*Corresponding author: mykolak@htaphotomask.com*

**We explore a new class of Distributed Feedback (DFB) structures that employ the recently-developed concept of Parity-Time (PT) symmetry in optics. We demonstrate that, based on PT-symmetric pure reflective volume gratings, vertical cavity surface-emitting lasers can be constructed. We provide a detailed analysis of the threshold conditions as well as the wavelength and angular spectral characteristics using the Kogelnik coupled-wave approximation, backed up by an exact solution of the Helmholtz equation. We demonstrate that such a PT-symmetric cavity can be configured to support one and only one longitudinal mode, leading to inherently single-mode lasing.**

*OCIS codes: (050.0050) Diffraction and gratings; (250.7270) Vertical emitting lasers; (140.3495) Laser, distributed feedback; (140.4480) Optical amplifiers; (160.3918) Metamaterials.*

The vertical light-emitting geometry of active optoelectronic devices such as vertical cavity surface-emitting lasers (VCSEL) and vertical external cavity surface emitting lasers (VECSEL) has revolutionized a wide range of high-tech applications, particularly optical fiber communication, optical digital recording (CD, DVD, and Blu-Ray), laser materials processing, biology and medicine, spectroscopy, imaging, and many others.

In conventional lasers a longer cavity produces multiple competing modes. However, recent studies on Parity-Time (PT) symmetric resonators [1-3] demonstrated intrinsically single-mode lasing regardless of the gain spectral bandwidth. Originally this was predicted theoretically in Distributed Bragg Reflectors with an embedded PT-symmetric grating inside [1] and in a cavity formed by two PT-symmetric gratings [2]. Later an intrinsically single-mode lasers was also explained theoretically and built experimentally [3] in a ring resonator with a built-in PT-symmetric grating. Our objective is to design a laser based on a pure reflective volume PT-symmetric grating that irradiates perpendicular to the grating slab, generating one and only one fundamental mode. This laser is not capable of supporting any higher-order mode.

In optics, PT-symmetric materials are artificial structures that combine gain and loss with refractive index modulation so as to mimic the complex PT-symmetric potential in quantum physics [4] where PT-symmetry was originally proposed as an alternative criterion for non-Hermitian Hamiltonians to possess a real eigenspectrum. In this Letter we extend our recent studies [5, 6] of a slab with a pure reflective PT-symmetric grating on configurations that support lasing. The proposed configuration is shown in Fig.1 (a) in a schematic form. It consists of a PT-symmetric grating on a substrate with a perfectly-reflecting mirror between them.

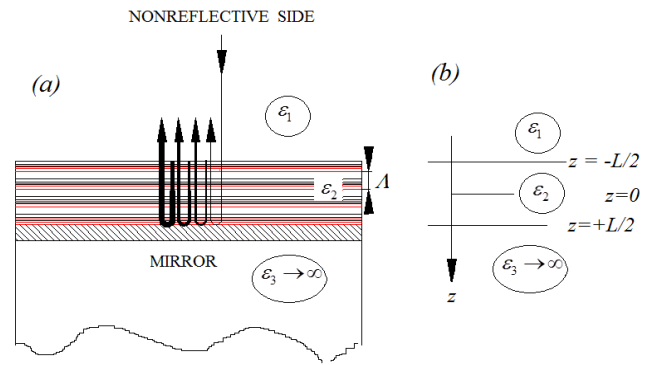

Fig. 1. Schematic view of the PT-symmetrical vertically emitting laser based on a planar purely reflective volume grating with index (black color fringes) and gain/loss (red color fringes) modulations.

This pure reflective grating is assumed to be composed of harmonic modulation of the relative dielectric permittivity and gain/loss modulation with the same spatial frequency shifted by a quarter of period $\Lambda/4$ with respect to each other. Its balanced state is achieved when the amplitudes of the index modulation and the gain/loss modulation are equal, i.e. $\Delta\varepsilon = c\mu\Delta\sigma/k_0$, where $\Delta\varepsilon$ is the amplitude of the relative permittivity distribution, $\Delta\sigma$ is the amplitude of the gain/loss periodic distribution, $\mu$ is the permeability of the medium, $\omega$ is the angular frequency of the wave, and $k_0 = \omega/c$ is the wave-vector in free space, related to the free-space wavelength $\lambda_0$ by $k_0 = 2\pi/\lambda_0$, and $c$ is the velocity of light *in vacuo*. Such a balanced PT-

symmetric grating can be represented by the following complex distribution of the dielectric constant:

$$\varepsilon(x,z) = \varepsilon_2\left(1 + (\xi/2)\exp(j(2\beta z + \phi))\right) \quad (1)$$

in the region from $z = -L/2$ to $z = +L/2$ where $\varepsilon_2$ is the average relative permittivity of the grating, $\xi = 2\Delta\varepsilon/\varepsilon_2$ and $\beta=\pi/\Lambda$. In Eq. (1) we have introduced a phase $\phi$, which corresponds to a possible shift in $z$.

In our recent publications [5, 6] we have developed techniques to accurately analyze diffraction of *H*-mode polarized light ($E_x=E_z=0$, $E_y \neq 0$) on such PT-symmetric gratings. The pure reflective PT-symmetric grating is very asymmetric in its diffraction; it has a diffractive side where incident light experiences strongly amplified back diffraction, whereas light incident on opposite (non-diffractive) side of the grating passes through the grating without any diffraction. For achieving lasing the non-diffractive side should be the side of incidence, with strong reflection from the opposite (diffractive) side provided, for example, by placing a highly-reflective metal layer there, as shown in Fig. 1(a). Only in this configuration is a positive feedback provided that can lead to lasing when a threshold is reached. Indeed, incident light under any incident angle (only normal incidence is shown in Fig. 1(a)) reaches the mirror without any significant reflection. After being reflected from the mirror the light is reflected back again toward the mirror, experiencing strong back-diffraction with amplification. This process is repeated again and again, generating more and more light back into free space, as shown schematically in Fig. 1(a) for normal incidence. It is important to point out that in the configuration described there is no need for a top reflector (mirror). The Fresnel reflection from the $\varepsilon_1/\varepsilon_2$ interface can affect the threshold, but it is not critical in achieving lasing.

Our objective is to design the PT-symmetric grating with the bottom reflector in such a way that this structure is able to generate at one and only one fundamental mode and is unable to lase at any high-order mode. This can be achieved by carefully engineering gain/loss and index modulations along with their phases. We modified the well-known Kogelnik coupled-wave theory to account for Fresnel reflection at the slab boundary when $\varepsilon_1 \neq \varepsilon_2$.

The electric field outside the slab and within the slab can be represented as a sum of forward (f) and backward (b) propagating waves (Fig. 1(b)):

$$\begin{aligned} E_1 &= \exp(jk_1 z) + R_L \exp(-jk_1 z) \\ E_2 &= E_f(z)\exp(jk_2 z) + E_b(z)\exp(-jk_2 z) \end{aligned} \quad (2)$$

where $k_1 = k_0(\varepsilon_1)^{1/2}$; $k_2 = k_0(\varepsilon_2)^{1/2}$. The boundary conditions for the electric field at the slab boundary ($z=-L/2$), namely continuity of $E$ and $E'$, lead to the transfer matrix for the boundary:

$$M^{(1,2)} = \frac{1}{2}\begin{pmatrix} (1+\gamma)\exp\left(j\frac{k_2-k_1}{2}L\right) & (1-\gamma)\exp\left(j\frac{k_1+k_2}{2}L\right) \\ (1-\gamma)\exp\left(-j\frac{k_2+k_1}{2}L\right) & (1+\gamma)\exp\left(j\frac{k_1-k_2}{2}L\right) \end{pmatrix} \quad (3)$$

where $\gamma = k_1/k_2 = (\varepsilon_1)^{1/2}/(\varepsilon_2)^{1/2}$.

The transfer matrix for the PT-symmetric grating in its non-diffractive orientation has the following form [8]:

$$M^{(PT)} = \begin{pmatrix} 1 & m_{12}^{(PT)} \\ 0 & 1 \end{pmatrix} \quad (4)$$

where

$$m_{12}^{(PT)} = jk_2 L \frac{\sin(\delta L)}{2\delta L} \xi \exp(j\phi) \quad (5)$$

and $\delta = k_2-\beta$ is the detuning from the Bragg condition.

The electric field amplitudes at the mirror ($z=+L/2$) can then be found from the following matrix equation:

$$\begin{pmatrix} E_f(L/2) \\ E_b(L/2) \end{pmatrix} = M^{(PT)} M^{(1,2)} \begin{pmatrix} 1 \\ R_L \end{pmatrix} \quad (6)$$

Using the reflection condition: $E_f(+L/2)\exp(jk_2L/2) + E_b(+L/2)\exp(-jk_2L/2)=0$ at a perfect mirror, we can obtain the final expression for the reflection coefficient:

$$R_L = -\frac{(1+\gamma)\exp(jk_2L) + (\exp(-jk_2L) + m_{12}^{(PT)})(1-\gamma)}{(1-\gamma)\exp(jk_2L) + (\exp(-jk_2L) + m_{12}^{(PT)})(1+\gamma)} e^{jk_1L} \quad (7)$$

The lasing threshold condition occurs when the denominator of Eq. (7) is zero. In general the phase $\phi$ in the expression for $m_{12}{}^{(PT)}$ could be arbitrary, but we consider here only two cases: **(i)** $\phi=0$, when $m_{12}{}^{(PT)}$ is purely imaginary, and **(ii)** $\phi=\pi/2$, when $m_{12}{}^{(PT)}$ is real. In the following discussion we will take the grating length $L$ to be an integral number of grating periods, i.e. $L=m\Lambda$.

**(i)**     $\phi=0$, $m_{12}{}^{(PT)}$ pure imaginary

The lasing condition is obtained by equating the real and imaginary parts of the denominator to zero.

$$\cos(k_2 L) = 0 \quad (8a)$$

$$\frac{2\gamma}{1+\gamma}\sin(k_2 L) = k_2 L \frac{\xi}{2} \frac{\sin(\delta L)}{\delta L} \quad (8b)$$

The real part of the denominator, Eq. (8a), gives us the frequency lasing condition, while the imaginary part, Eq. (8b), provides us with the threshold value. The real part, Eq. (8a), is equal to zero only when $k_2L = (i+1/2)\pi$, which, on writing $i=m+r$, is equivalent to $k_2=\pi/\Lambda+(r+1/2)\pi/L$. Then the detuning is $\delta = k_2-\beta = (r+1/2)\pi/L$, where $r = 0, \pm 1, \pm 2, ...$. In other words, the non-Hermitian vertical-emitting laser with a PT-symmetric grating of an integral number of periods can lase at discrete wave-numbers away from the Bragg condition ($\delta=0$), i.e. $\delta=\pm\pi/2L; \pm 3\pi/2L; \pm 5\pi/2L, ...$ with the fundamental lasing mode occurring at $\delta=\pm\pi/2L$.

The threshold value for the PT-symmetric grating strength can be found from Eq. (8b) under the condition that $\sin(k_2L) = (-1)^{m+r}$:

$$\xi_{th}^{(r)} = (-1)^m \frac{2\gamma}{1+\gamma} \frac{(2r+1)}{(m+r+1/2)} \quad (9)$$

The threshold should have a positive value, so if $m$ is odd, $r$ must be negative, leading to a "blue" shift of the lasing wavelength, $\delta=-\pi/2L; -3\pi/2L; -5\pi/2L, ...$, while if $m$ is even we get a "red" shift, $\delta=\pi/2L; 3\pi/2L; 5\pi/2L, ...$. As a result, for the even-period gratings the lowest threshold lasing starts when $r=0$ with the shift $\delta=\pi/2L$ or $\Delta v=c/4n_2L$ from the Bragg frequency. The thresholds have slightly different expressions for gratings with even and odd numbers of periods, but in fact the PT-symmetric gratings with $m=72$ periods (even) and with $m=73$ periods (odd) have the same threshold $2\gamma/(72.5(1+\gamma))$, but with opposite shifts with respect to the Bragg wavelength ($\lambda_B=2\Lambda/(\varepsilon_2)^{1/2}=632.071$ nm). The frequency spacing to the next lasing mode is $c/(2(\varepsilon_2)^{1/2}L)$. This frequency spacing is equal to the longitudinal mode spacing of a traditional two-mirror cavity of the same length $L$ as the PT-symmetric grating. Clearly, such a structure does not produce the desirable single-mode operation.

**(ii) $\phi=\pi/2$, $m_{12}^{(PT)}$ real**

It is easy to notice that the mode spacing in k-space $\Delta k = \delta = \pi/2L$ corresponds exactly to the spacing between the zeros of the sinc function that is a specific feature of this PT-symmetric grating [8]. If the lasing modes can be shifted so that the fundamental mode occurs at $\delta=0$, the structure will support only a single lasing mode coinciding with the main lobe of the sinc function, as is illustrated in Fig. 2(b)

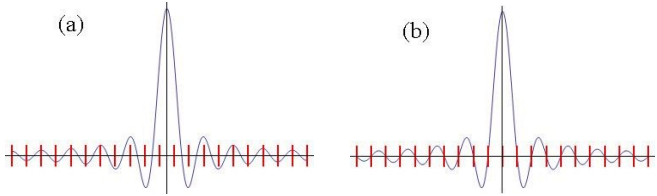

Fig. 2. Possible lasing modes for (a) $\phi=0$, (b) $\phi=\pi$. The red bars indicate the possible values of $k_2L$, while the blue curve is the sinc function of Eq. (5).

This can be achieved by taking $\phi=\pi/2$, i.e. phase-shifting the PT-symmetric grating with respect to the mirror at the bottom and the $\varepsilon_1/\varepsilon_2$ interface by a quarter period, $\Lambda/4$. This results in $m_{12}^{(PT)}$ becoming real, and corresponds to the following complex dielectric constant modulation:

$$\varepsilon(x,z) = \varepsilon_2(1 + j(\xi/2)\exp(2j\beta z)) \quad (10)$$

By engineering the complex modulation in this way the threshold condition takes the following form:

$$\sin(k_2 L) = 0 \quad (11a)$$

$$\frac{2}{1+\gamma}\cos(k_2 L) = k_2 L \frac{\xi}{2}\frac{\sin(\delta L)}{\delta L} \quad (11b)$$

As a result of the change of phase, it is now the imaginary part, Eq. (11a), that determines the admissible values of $k_2$, namely $k_2L=s\pi$, which, upon writing $s=m+r$, is equivalent to $\delta=k_2-\beta=r\pi/L$, where $r=0,\pm1,\pm2,\ldots$.

The phase condition is satisfied for all these $r$ values. However, Eq. (11b) has a solution only for the fundamental mode, $r=0$, because sinc($\delta L$) is zero for all higher order modes. Thus the PT-symmetric cavity in this configuration supports only the fundamental mode. The cavity becomes completely transparent at the wave-numbers for any higher order modes, and such modes simply cannot be amplified. The threshold condition is

$$\xi_{th}^{(0,m\Lambda)} = (-1)^m \frac{4}{\pi m(1+\gamma)} \quad (12)$$

with $m$ even. If instead we took $\phi=-\pi/2$, lasing would occur for $m$ odd.

Bearing in mind that the coupling coefficient of the PT-symmetric grating is $\kappa = \pi(2\Delta\varepsilon)/\lambda$, the threshold condition can be presented in the more traditional form: $\kappa L=2/(\gamma+1)$, or in the case of zero Fresnel reflection $\kappa L=1$, which exactly coincides with the threshold value in the case of two concatenated PT-symmetric gratings with their non-diffractive sides oriented outwards from the cavity [2]. This threshold value, $\kappa L=1$, is $2/\pi$ times lower that the threshold value for a traditional DFB laser according to Kogelnik and Shank [9]. The constructive contribution of the Fresnel reflection can be used to reduce the threshold even lower, based on the factor $2/(\gamma+1)$ when $\varepsilon_1 > \varepsilon_2$.

The external dielectric constant, $\varepsilon_1$, strongly influences the threshold condition through Fresnel reflection. As already mentioned, there is no need for an external reflector in the proposed structure because the PT-symmetric grating provides this reflectivity and the reflective feedback. However, the more light is reflected back into the cavity by the interface in addition to the light reflected by the grating, the more power is involved in the positive feedback and the lower the threshold should be if these two reflected light waves interfere constructively. This is exactly what occurs in the case $\varepsilon_2 > \varepsilon_1$. The higher the $\varepsilon_2/\varepsilon_1$ ratio, the lower the threshold compared with the case of zero Fresnel reflection, when $\varepsilon_2 = \varepsilon_1$, and $\xi_{th}^{(0)} = 2/(\pi m)$. On the other hand, when $\varepsilon_1 > \varepsilon_2$, the Fresnel reflected light and the grating reflected light interfere destructively, so the threshold increases. In Fig. 3 the red curve shows the threshold dependence on the external dielectric constant for $\varepsilon_2 = 2.4$ and $m = 72$ based on Eq. (12).

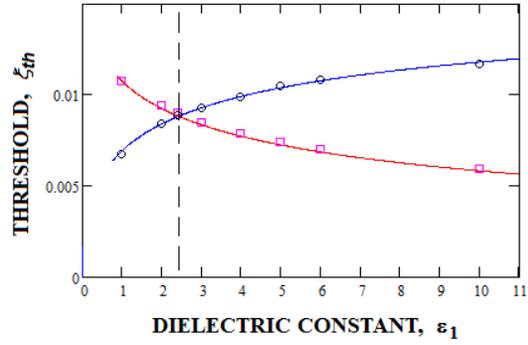

Fig. 3. The threshold value as a function of the external dielectric constant; the red and blue curves are based on the approximate Kogelnik method (Eqs. (12) and (13) respectively) and the magenta diamonds and black circles are numerically calculated values based on the exact Bessel function method [6]. The vertical dashed line shows the threshold for zero Fresnel reflection, when $\varepsilon_1 = \varepsilon_2$.

It is possible to adjust the phase of the PT-symmetric grating within the non-Hermitian vertical cavity laser in such a way that its interaction with the $\varepsilon_1/\varepsilon_2$ interface will be opposite to the situation shown in Fig. 3 by the red curve. This follows from Eq. (8) when, instead of consisting of a whole number of grating periods, the PT-symmetric grating has a half-integral number of periods, i.e. $L = (m+1/2)\Lambda$. In that case it is easy to show that such a configuration also provides intrinsically single-mode lasing at the Bragg wavelength with the threshold condition:

$$\xi_{th}^{(0,(m+1/2)\Lambda)} = \frac{2\gamma}{\pi(1+\gamma)}\frac{2}{(m+1/2)} \quad (13)$$

As a result of the change in phase, the Fresnel contribution now interferes constructively with the PT-symmetric grating reflection for $\varepsilon_1 < \varepsilon_2$ and destructively for $\varepsilon_1 > \varepsilon_2$, as shown by the blue curve in Fig. 3, with a few calculated values from the exact method. For zero Fresnel reflection ($\varepsilon_1 = \varepsilon_2$) the threshold is clearly the same for the both configurations. The blue curve in Fig. 3 shows the threshold dependence on the external dielectric constant for $\varepsilon_2 = 2.4$ and $m = 72$ based on Eq. (13).

The results obtained by the modified Kogelnik method can be substantiated by exact calculations based on an analytical solution of the scalar Helmholtz equation using the modified Bessel functions. The solutions previously obtained for the one-dimensional PT-symmetric grating with a periodic perturbation proportional to exp($2j\beta z$) [7] have been generalized [6] to cover the case of different refractive indices on either side of the grating as well as to non-normal incidence. The solutions were used to explore the transmission and refection characteristics of the grating as a function of incident angle in a variety of situations. These solutions can be used to find the reflectivity from the non-diffractive side of the slab when its opposite surface is a perfectly reflecting mirror by taking $\varepsilon_3 \to \infty$.

We are interested in vertical single-mode operation, so the grating period should be adjusted to produce diffraction perpendicular to the slab rather than lasing at an angle to the substrate. For example, if we take the PT-grating with period $\Lambda$= 0.204 µm and $\lambda$=632.07 nm, then the diffraction occurs at the Bragg angle $\theta_B$ = 0.85⁰. The spectral characteristics of such a structure are shown in Figs. 4(a) and 4(c) at normal incidence, $\theta$ =0, for the coupling strength below the threshold, $\xi$=0.005, and practically at the threshold $\xi =\xi_{th}$ =0.00685, respectively. The dashed blue vertical line shows the lasing wavelength, which coincides with the Bragg wavelength in Fig. 4(c).

The proposed non-Hermitian surface emitting laser also exhibits a very nice angular spectrum, as shown in Fig. 4(b) below the threshold and in Fig. 4(d) at the threshold for the lasing wavelength ($\lambda_{th} = 632.07$ nm), with a divergence of less than 0.3⁰ in the internal refraction angle, $\theta$.

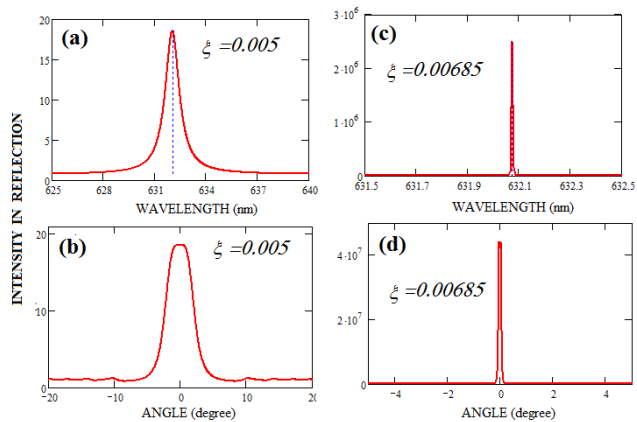

Fig. 4. Reflective wavelength (a, c) and angular (b, d) spectra for $\varepsilon_1$ =6; $\varepsilon_2$ = 2.4; $\varepsilon_3$ = ∞; $d$=72$\Lambda$; $\Lambda$=0.204 µm at $\xi = 0.005$ (a, b) and at the threshold condition: $\xi = \xi_{th} = 0.00685$ (c, d).

It is important to mention that, for these parameters at least, the modified Kogelnik approach based on various approximations turns out to be very accurate in providing the threshold values. For visual comparison a few threshold values calculated using the exact method are plotted in Fig. 3 as the boxes or circles near the red and blue curves, respectively. As we see, the results from the two different methods agree very well.

In conclusion, we have proposed and simulated a surface-emitting laser based on a *PT*-symmetric DFB structure. The vertical cavity is formed by the balanced, pure reflective volume *PT*-symmetric grating. The grating is oriented towards the mirror by its diffractive (reflective side), and amplified emission occurs towards the grating non-diffractive (non-reflective) side into free space perpendicular to the substrate surface or at an angle to the surface if required. We used two methods to analyze the threshold condition, the mode lasing pattern and the spectral and angular emission characteristics of the proposed non-Hermitian laser.

Our analysis shows that the proposed structure exhibits unusual lasing characteristics. The structure can be configured in such a way that it becomes intrinsically single mode regardless of the gain spectral bandwidth. The influence of the Fresnel reflection from the boundary between the grating slab and the superstrate was analyzed, and we demonstrated that this reflection can be used to reduce the threshold by more than 30%.

The proposed structure paves a novel route for designing a new class of multifunctional optical active devices with completely nonsymmetrical optical responses and new mechanisms for light manipulation.